\documentclass[aps,prc,preprint,showpacs,superscriptaddress,showkeys,tightenlines]{revtex4}
\usepackage{epsfig,rotating}

\begin{document}
\title{Momentum distributions in stripping reactions of radioactive projectiles at
intermediate energies}
\author{C.\,A.~Bertulani}
\affiliation{Department of Physics and Astronomy, Michigan State
University, East Lansing, Michigan 48824}
\author{P.\,G.~Hansen}
\affiliation{Department of Physics and Astronomy, Michigan State
University, East Lansing, Michigan 48824}
   \affiliation{National Superconducting Cyclotron Laboratory,
     Michigan State University,
     East Lansing, Michigan 48824}
\email{hansen@nscl.msu.edu}
\date{\today}

\begin{abstract}
The theory of one-nucleon removal in the stripping reaction
(inelastic breakup) on a light target is extended to cover
two-dimensional momentum distributions of the reaction residues
with the use of realistic profile functions for the core-target
and nucleon-target interactions. Examples of the calculated
projected parallel- and transverse momentum distributions are
given. The transverse momentum distributions, projections on a
Cartesian axis perpendicular to the beam direction, show an
interesting intermingling of the stripping reaction with elastic
scattering of the reaction residue on the target. We also obtain
doubly-differential distributions of the cross section on the
parallel- and transverse-momentum variables. The distributions
depend strongly on the value of the magnetic quantum number $m$.
They will be of importance for evaluating acceptance corrections
in experiments, and they lead to alignment with the possibility of
anisotropic emission of subsequent gamma rays, an interesting
spectroscopic tool. Experimental data for proton stripping of
$^{8}$B agree with our calculations.
\end{abstract}


\maketitle

\section{Introduction}
\label{intr} Single-nucleon knockout reactions with heavy ions, at
intermediate energies and in inverse kinematics, have become a
specific and quantitative tool for studying single-particle
occupancies and correlation effects in the nuclear shell model,
see the recent review \cite{han03}. The high sensitivity of the
method has allowed measurements on rare radioactive species
available in intensities of less than one atom per second for the
incident beam. The experiments observe reactions in which fast,
mass $A$, projectiles with laboratory momentum $\mathbf{k}_{A} =
\mathbf{P}_{A} / \hbar$ collide peripherally with a light nuclear
target, typically $^{9}$Be, producing residues with mass $(A-1)$,
in the following referred to as the core ($c$) of the assumed
two-body system of core plus nucleon. In the laboratory system the
momentum transferred to the core is
\begin{equation}
\mathbf{k}_{c,lab}=\frac{A-1}{A}\mathbf{k}_{A}-\mathbf{k}_{A-1}. \label{kbal}%
\end{equation}
The final state of the target and that of the struck nucleon are
not observed, but instead the energy of the final state of the
residue can be identified by measuring coincidences with decay
gamma-rays emitted in flight. Referred to the center-of-mass
system of the projectile, the transferred momentum is
$\mathbf{k}_{c}$. In the sudden approximation and for the
stripping reaction, defined below, this must equal the momentum of
the struck nucleon before the collision. The measured partial
cross sections to individual final levels provide spectroscopic
factors for the individual angular-momentum components $j$. In
complete analogy to the use of angular distributions in transfer
reactions, the orbital angular momentum $l$ is in the knockout
reactions revealed by the distributions of the quantity
$\mathbf{k}_{c}$. These distributions are the subject of the
present paper.

The early interest in momentum distributions came from studies of
nuclear halo states, for which the narrow momentum distributions
in a qualitative way revealed the large spatial extension of the
halo wave function. It was pointed out by Bertulani and McVoy
\cite{ber92} that the longitudinal component of the momentum
(taken along the beam or $z$ direction) gave the most accurate
information on the intrinsic properties of the halo and that it
was insensitive to details of the collision and the size of the
target. In contrast to this, the transverse distributions of the
core are significantly broadened by diffractive effects and by
Coulomb scattering. For experiments that observe the nucleon
produced in elastic breakup, the transverse momentum is entirely
dominated by diffractive effects, as illustrated \cite{ann94} by
the angular distribution of the neutrons from the reaction
$^{9}$Be($^{11}$Be,$^{10}$Be+n)X. In this case, the width of the
transverse momentum distribution reflects essentially the size of
the target. Experiments and theory for reactions of neutron halos
have bee reviewed in ref. \cite{han95}. It was found that to
understand the measured longitudinal momentum distributions it is
necessary to take into account that a heavy-ion knockout reaction,
being surface-dominated, can only sample the external part of the
nucleon wave function. The magnitude of the reaction cross section
is determined by the part of the wave function that is accessed,
and the shape of the momentum distribution reflects the momentum
content in this part. Calculations \cite{han95,han96,esb96} based
on a sharp-surface strong-absorption (\textquotedblleft
black-disk\textquotedblright) model could account for the observed
longitudinal momentum distributions and also, approximately, for
the absolute cross sections. This approach is confirmed in the,
more accurate, present work, which extends the theory to include
the general dependence of the differential cross section on the
momentum vector.

It is essential to note that the cross section for the production
of a given final state of the residue has two contributions. The
most important of the two, commonly referred to as stripping or
inelastic breakup, represents all events in which the removed
nucleon reacts with and excites the target from its ground state.
The second component, called diffractive or elastic breakup,
represents the dissociation of the nucleon from the residue
through their two-body interactions with the target, each being at
most elastically scattered. These events result in the ejected
nucleon being present in the forward beam with essentially beam
velocity, and the target remaining in its ground state. These
processes lead to different final states, they are incoherent, and
their cross sections must be added in measurements where only the
residue is observed. General expressions for the total and
differential cross sections for the two components have been given
by Hencken, Bertsch and Esbensen \cite{hen96}.

In a subsequent development, the knockout method was extended to
non-halo states \cite{nav98,tos99,tos01,nav00,aum00,mad01,end03}.
For these, involving more deeply-bound nucleons, the one-nucleon
stripping cross sections are much smaller than the free-nucleon
reaction cross section on the same target; a ratio that gives a
measure of how much the nucleon wave function is \textquotedblleft
shielded\textquotedblright\ from the target by the bulk of the
core. This required a more elaborate theoretical treatment based
on the elastic ${S}$-matrices ${S}_{c}$ and ${S}_{n}$
\cite{alk96,tos97} of the core and nucleon. For a general review
of applications of this technique see ref.\,\cite{han03}, which
shows that very accurate theoretical single-particle cross
sections are obtained in this way. However, the longitudinal
momentum distributions have continued being calculated in the
black-disk approximation. This simplification has been permissible
because the assumed sharp surface of the target generates only
transverse momentum components, which are integrated out in the
final result.

In the present paper we treat the three-dimensional momentum
distribution of the core in stripping reactions at the same level
of approximation as the single-particle cross sections in
Tostevin's calculations \cite{tos99,tos01}. Using ${S}$-matrices
from this work, we obtain identical single-particle cross sections
after integration over momentum and summation over the $m_{l}$
substates. The results (i) test the reliability of the previously
used sharp-surface approximation, and (ii) demonstrate that the
transverse momentum distributions are quantitatively and even
qualitatively different from the parallel momentum distributions,
and (iii) can serve to extract angular-momentum information from
the angular distributions of the residues. Finally, (iv) the
results are of importance for calculating acceptance corrections
in experiments, and we evaluate the correlations between
longitudinal and transverse components. After a presentation of
the main features of the theoretical calculations, we give
examples of the distributions obtained for different values of the
nucleon separation energy $S_{n}$ and the orbital angular momentum
$l$. Some essential numerical details are presented in an
appendix.

\section{Calculation of the differential stripping cross section}
\label{calc} In this section we summarize the equations used in
the calculation of stripping cross sections. The numerical
calculations were done by using a Gaussian expansion method which
is described in the appendix. We also show the equations used for
the elastic scattering cross section, which will be important to
interpret the numerical results obtained for the stripping cross
sections, discussed in section 3.
\begin{figure}[tbp]
\begin{center}
\includegraphics[scale=0.4,angle=0]{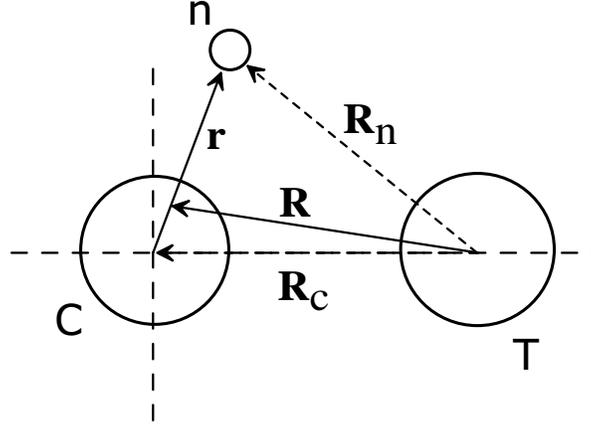}
\end{center}
\caption{Coordinates used in text.} \label{coord}
\end{figure}

Following ref. \cite{hen96}, the cross section for the stripping
reaction $(c+n)+A\longrightarrow c+X$, where $c$ corresponds to a
specified final state of the core, is given by%
\begin{equation}
\frac{d\sigma_{\mathrm{str}}}{d^{3}k_{c}}=\frac{1}{\left(  2\pi\right)  ^{3}%
}\frac{1}{2l+1}\sum_{m}\int d^{2}b_{n}\left[  1-\left\vert S_{n}\left(
b_{n}\right)  \right\vert ^{2}\right]  \left\vert \int d^{3}r\ e^{-i\mathbf{k}%
_{c}\mathbf{.r}}S_{c}\left(  b_{c}\right)  \psi_{lm}\left(  \mathbf{r}\right)
\right\vert ^{2}, \label{sknock}%
\end{equation}
and where $\mathbf{r\equiv}\left(
\mathbf{\mbox{\boldmath$\rho$}},z,\phi\right)
=\mathbf{R}_{n}-\mathbf{R}_{c}$, so that
\begin{eqnarray}
b_{c}  &  =\left\vert \mathbf{\mbox{\boldmath$\rho$}}-\mathbf{b}%
_{n}\right\vert =\sqrt{\rho^{2}+b_{n}^{2}-2\rho\ b_{n}\cos\left(  \phi
-\phi_{n}\right)  }\nonumber\\
&  =\sqrt{r^{2}\sin^{2}\theta+b_{n}^{2}-2r\sin\theta\ b_{n}\cos\left(
\phi-\phi_{n}\right).  }%
\end{eqnarray}
with the symbols ${\bf b}_c$ and ${\bf b}_n$ denoting
two-dimensional vectors, which are the respective transverse
components of ${\bf R}_c$ and ${\bf R}_n$ (see figure \ref{coord}).
$S_{c}$ ($S_{n}$) are the S-matrices for the core+target and the
neutron (or proton)+target scattering.

The single-particle bound state wave functions for the subsystem
$(c+n)$, i.e. $\psi_{lm}\left(  \mathbf{r}\right)  $, in eq.
(\ref{sknock})\ are specified by\ $\psi_{lm}\left(
\mathbf{r}\right)  =R_{l}\left(  r\right)  Y_{lm}\left(
\widehat{\mathbf{r}}\right)  $, where $R_{l}\left(  r\right)  $ is
the radial wave function. It is not necessary to specify the total
single-particle angular momentum $j$, since the assumed
interaction is spin-independent, and the depth of the
single-particle potential well is adjusted to reproduce the
effective nucleon binding energy.

The cross sections for the longitudinal momentum distributions are obtained by
integrating eq. (\ref{sknock}) over the transverse component of $\mathbf{k}%
_{c}$, i.e. over $\mathbf{k}_{c}^{\perp},$ and using%
\begin{equation}
\int d^{2}\mathbf{k}_{c}^{\perp}\ \exp\left[  -i\mathbf{k}_{c}%
\mathbf{.(\mbox{\boldmath$\rho$}-\mbox{\boldmath$\rho$}}^{\prime})\right]
=\left(  2\pi\right)  ^{2}\delta\left(
\mathbf{\mbox{\boldmath$\rho$}-\mbox{\boldmath$\rho$}}^{\prime}\right)  \ .
\label{DiracT}%
\end{equation}

One gets%
\begin{eqnarray}
\frac{d\sigma_{\mathrm{str}}}{dk_z}  &  =& \frac{1}{2\pi}\frac{1}%
{2l+1}\sum_{m}\int_{0}^{\infty}d^{2}b_{n}\ \left[  1-\left\vert S_{n}\left(
b_{n}\right)  \right\vert ^{2}\right]  \ \ \int_{0}^{\infty}d^{2}%
\rho\ \left\vert S_{c}\left(  b_{n}\right)  \right\vert ^{2}\nonumber\\
&  \times&\left\vert \int_{-\infty}^{\infty}dz\ \exp\left[
-ik_{z}z\right]  \psi_{lm}\left(  \mathbf{r}\right) \right\vert
^{2}\ ,
\label{strL}%
\end{eqnarray}
where $k_z$ represents the longitudinal component of ${\bf k}_c$.

For the transverse momentum distribution in cylindrical
coordinates $k_\bot=\sqrt{k_{x}^{2}+k_{y}^{2}}$, one uses in eq.
(\ref{sknock})
\begin{equation}
\int_{-\infty}^{\infty}dk_z\ \exp\left[  -ik_z(z-z^{\prime
})\right]  =2\pi\delta\left(  z-z^{\prime}\right)  \ , \label{DiracZ}%
\end{equation}
and the result is
\begin{eqnarray}
\frac{d\sigma_{\mathrm{str}}}{d^{2}k_\bot}  &  =&
\frac{1}{2\pi}\frac {1}{2l+1}\ \int_{0}^{\infty}d^{2}b_{n}\ \left[
1-\left\vert S_{n}\left(
b_{n}\right)  \right\vert ^{2}\right] \nonumber\\
&  \times& \sum_{m,\ p}\ \int_{-\infty}^{\infty}dz\ \left\vert \int d^{2}%
\rho\ \exp\left(  -i\mathbf{k}_{c}^{\perp}\mathbf{.\mbox{\boldmath$\rho$}}%
\right)  S_{c}\left(  b_{n}\right)  \psi_{lm}\left(  \mathbf{r}\right)
\right\vert ^{2}. \label{strT}%
\end{eqnarray}

Sometimes it is convenient to describe the transverse momentum
distributions in terms of the projection onto one of the Cartesian
components of the transverse momentum. This can be
obtained directly from eq. (\ref{strT}), i.e.%
\begin{equation}
\frac{d\sigma_{\mathrm{str}}}{dk_{y}}=\int dk_{x}\ \frac{d\sigma
_{\mathrm{str}}}{d^{2}k_\bot}\left(  k_{x},k_{y}\right)  \ . \label{sigtx}%
\end{equation}

The total stripping cross section can be obtained by integrating
either eq. (\ref{strL}) or eq. (\ref{strT}). For example, from eq.
(\ref{strL}), using eq.
(\ref{DiracZ}), one obtains%
\begin{eqnarray}
\sigma_{\mathrm{str}}  &  =& \frac{2\pi}{2l+1}\int_{0}^{\infty}db_{n}%
\ b_{n}\ \left[  1-\left\vert S_{n}\left(  b_{n}\right)  \right\vert
^{2}\right]  \ \nonumber\\
&  \times&\int d^{3}r\ \left\vert S_{c}\left(
\sqrt{r^{2}\sin^{2}\theta +b_{n}^{2}-2r\sin\theta\
b_{n}\cos\phi}\right)  \right\vert ^{2}\ \sum _{m}\left\vert
\psi_{lm}\left(  \mathbf{r}\right)  \right\vert ^{2}\ ,
\end{eqnarray}
which is the same as eq. (12) of ref. \cite{hen96}.

In the Appendix A we show how the integrals in eqs. (\ref{strL})
and (\ref{strT}) can be evaluated numerically with use of Gaussian
expansions of the core/nucleon+target S-matrices.

The S-matrices have been obtained using the eikonal approximation
for the wave functions. In this approximation the outgoing wave of
a fragment, with wave number ${\bf k}$, is given by
\begin{equation}
\left\langle{\bf r}|\Psi^{(+)}_{\bf k}\right\rangle = \exp\left\{
i {\bf k}\cdot {\bf r} +{i \over \hbar v}\int_z^\infty dz'
U_{opt}(r')\right\} \ . \label{neweq}
\end{equation}
The overlap of the incoming and outgoing wave function becomes
\begin{equation}
\left\langle\Psi^{(-)}_{\bf k}|\Psi^{(+)}_{\bf
k'}\right\rangle=S\left( b\right)
\exp\left(  i\mathbf{q.r}\right)  \ , \label{intro2}%
\end{equation}
where ${\bf q}={\bf k}'-{\bf k}$ is the momentum transfer and
$S\left( b\right) $ is the scattering matrix given by
\begin{equation}
S\left(  b\right)  =\exp\left[  i\chi(b)\right]  ,\ \ \ \ \ \ \ \ \text{with}%
\ \ \ \ \ \ \ \ \chi(b)=-{\frac{1}{\hbar
{v}}}\int_{-\infty}^{\infty
}dz\ U_{opt}(r)\ , \label{intro3}%
\end{equation}
and $U_{opt}(\mathbf{r})$ is the appropriate optical potential for the
core+target and the neutron (or proton)+target scattering. In equation
(\ref{intro3}) $\chi(b)$ is the eikonal phase, and $r=\sqrt{b^{2}%
+z^{2}}$, where $b$ is often interpreted as the impact parameter.
This interpretation arises from a comparison of the results
obtained with eikonal wavefunctions with those obtained with
classical particles colliding at a given impact parameter $b$
\cite{BD04}. Nonetheless, the eikonal wavefunction is a quantum
scattering state and $b$ is the transverse coordinate associated
to it. Thus wave-mechanical effects, like smearing and
interference, are accounted for properly.

In the optical limit of the Glauber theory, the eikonal phase is
obtained from the nuclear ground state densities and the
nucleon-nucleon cross sections by the relation \cite{BD04}
\begin{equation}
\chi(b)=\int_{0}^{\infty}dq\ q\ \rho_{p}\left(  q\right)
\rho_{t}\left(
q\right)  f_{NN}\left(  q\right)  J_{0}\left(  qb\right)  \ , \label{eikphase}%
\end{equation}
where $J_{0}$ is the Bessel function of order zero,
$\rho_{p,t}\left( q\right) $ is the Fourier transform of the
nuclear densities of the projectile and target, and $f_{NN}\left(
q\right)  $ is the high-energy nucleon-nucleon scattering
amplitude at forward angles, which can be parametrized
by \cite{ray79}%
\begin{equation}
f_{NN}\left(  q\right)  =\frac{k_{NN}}{4\pi}\sigma_{NN}\left(  i
+\alpha _{NN}\right)  \exp\left(  -\beta_{NN}q^{2}\right)  \ .
\end{equation}

In this equation $\sigma_{NN}$, $\alpha_{NN},$ and $\beta_{NN}$
are parameters which fit the high-energy nucleon-nucleon
scattering at forward angles. In eq. (\ref{eikphase}) the
quantities $\rho_{p}\left( q\right) $ and $\ \rho_{t}\left(
q\right)  $ are calculated from the radial density distributions,
usually \cite{han03} taken to be of Gaussian shapes for light
nuclei, and of Fermi shapes for heavier nuclei with parameters
taken from experiment. For cases where more accuracy is needed, it
is possible to take the density distributions from Hartree-Fock
calculations, as has been done in some recent work. This
demonstrates that the theoretical uncertainty on the integral
single-particle cross section ({\it i.e.} for a spectroscopic
factor of unity) is of the order of 5\% for a halo state
\cite{ter04} and 15\% for a very deeply bound $l=2$ state
($S_n=22$ MeV) \cite{gad04a}. The precise choice of input
parameters influences the absolute spectroscopic factors, which
have been the subject of many previous papers
\cite{han03,nav98,nav00,aum00,mad01,end03,ter04,gad04a} to which
we refer for numerical details. However, it means little for the
shapes of the momentum distributions which are the focus of the
present work, and it will not be discussed further here.

For the Coulomb part of the optical potential the integral in eq.
(\ref{intro3}) diverges. One solves this by using
$\chi=\chi_{N}+\chi_{C}$, where $\chi_{N}$ is given by  eq.
(\ref{intro3}) without the Coulomb potential and writing the
Coulomb eikonal phase, $\chi_{C}$ as
\begin{equation}
\chi_{C}(b)=2\eta\ln(kb)\ , \label{intro4}%
\end{equation}
where $\eta=Z_{p}Z_{t}e^{2}/\hbar v$, $Z_{p}$ and $Z_{t}$ are the
charges of projectile and target, respectively, $v$ is their
relative velocity, $k$ their wavenumber in the center of mass
system. Eq. (\ref{intro4}) reproduces the exact Coulomb scattering
amplitude when used in the calculation of the elastic scattering
with the eikonal approximation \cite{BD04}:
\begin{equation}
f_{C}(\theta)={\frac{Z_{p}Z_{t}e^{2}}{2\mu v^{2}\ \sin^{2}(\theta/2)}}%
\ \exp\left\{  -i\eta\ \ln\left[  \sin^{2}(\theta/2)\right]  +i\pi+2i\phi
_{0}\right\}  \label{fctheta}%
\end{equation}
where $\phi_{0}=arg\Gamma(1+i\eta/2)$. This is convenient for the
numerical calculations since, as shown below, the elastic
scattering amplitude can be written with the separate contribution
of the Coulomb scattering amplitude included. Then, the remaining
integral (the second term on the r.h.s. of eq. (\ref{simplif})
below) converges rapidly for the scattering at forward angles.

Although the Coulomb phase in eq. (\ref{intro4}) diverges at
$b=0$, this does not pose a real problem, since the strong
absorption suppresses the scattering at small impact parameters.
It is also easy to correct this expression to account for the
finite charge distribution of the nucleus. For example, assuming a
uniform charge distribution with radius $R$ the Coulomb phase
becomes
\begin{eqnarray}
\chi_{C}(b)  &  =2\eta\ \left\{  \Theta(b-R)\ \ln(kb)+\Theta(R-b)\left[
\ln(kR)+\ln(1+\sqrt{1-b^{2}/R^{2}})\right.  \right. \nonumber\\
&  \left.  \left.  -\sqrt{1-b^{2}/R^{2}}-{\frac{1}{3}}(1-b^{2}/R^{2}%
)^{3/2}\right]  \right\}  \ , \label{chico_0}%
\end{eqnarray}
where $\Theta$ is the step function. This expression is finite for
$b=0$, contrary to eq. (\ref{intro4}). If one assumes a Gaussian
distribution of charge with radius $R$, appropriate for light
nuclei, the Coulomb phase becomes
\begin{equation}
\chi_{C}(b)=2\eta\ \left[  \ln(kb)+{\frac{1}{2}}E_{1}(b^{2}/R^{2})\right]  \ ,
\label{chico2}%
\end{equation}
where the error function $E_{1}$ is defined as
\begin{equation}
E_{1}(x)=\int_{x}^{\infty}{\frac{e^{-t}}{t}}\ dt\ . \label{chico3}%
\end{equation}
This phase also converges as $b\rightarrow0$. In eq.
(\ref{chico_0}) $R=R_p+R_t$, while in eq. (\ref{chico2})
$R=\sqrt{R^2_p+R^2_t}$, where $R_p$ and $R_t$ are the respective
projectile and target radius. The cost of using the expressions
(\ref{chico_0}) and (\ref{chico2}) is that the Coulomb scattering
amplitude becomes more complicated than (\ref{fctheta}). Moreover,
we have numerically attested that the elastic and inelastic
scattering cross sections change very little by using eqs.
(\ref{chico_0}) or (\ref{chico2}), instead of eq. (\ref{intro4}).

In calculations involving stripping, the final state Coulomb
interaction between the core and the target is taken into account
by using the eikonal-Coulomb phase shift of (\ref{intro4}) in the
calculation of $S_{c}$. However in the calculation of diffraction
dissociation both $S_{c}$ and $S_{n}$ are calculated using the
eikonal-Coulomb phase shift of (\ref{intro4}).
\begin{figure}[tbp]
\includegraphics[scale=0.65,angle=0]{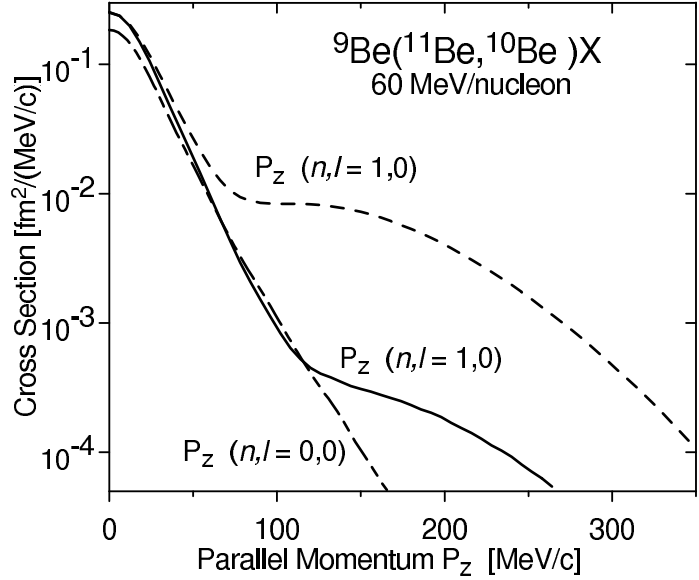}
\caption{ \label{be11z} Longitudinal momentum distribution for the
residue in the $^{9}$Be($^{11}$Be,$^{10}$Be$_{gs}$)X reaction at
60 MeV/nucleon as studied in the experiment of Aumann {\textit et
al.} \cite{aum00}. The dashed curve is the cross section
calculated in the transparent limit, re-scaled by a factor of
0.535. It has a high-energy component arising from the inner lobe
of the wave function. The full drawn curve is the result of the
present work. The black disk calculation (not shown) is
essentially indistinguishable from this, except for the
high-momentum tail. The dot-dashed curve shows that the weak
shoulder is not present with an assumed nodeless 0$s$ wave
function with the same binding energy (0.504 keV).}
\end{figure}

The calculation of elastic scattering amplitudes using eikonal
wave functions,
eq. (\ref{neweq}), is very simple. They are given by \cite{BD04}%
\begin{equation}
f_{el}(\theta)=ik\ \int_{0}^{\infty}db\ b\ J_{0}(qb)\ \left\{  1-\exp\left[
i\chi(b)\right]  \right\}  \ , \label{elast3}%
\end{equation}
where $q=2k\sin(\theta/2)$, and $\theta$ is the scattering angle. The elastic
scattering cross section is ${d\sigma_{el}/}d\Omega=\left\vert f_{el}%
(\theta)\right\vert ^{2}$. For numerical purposes, it is
convenient to make use of the analytical formula, eq.
(\ref{fctheta}), for the Coulomb scattering amplitude. Thus, if
one adds and subtracts the Coulomb amplitude, $f_{C}(\theta)$ in
eq. (\ref{elast3}), one gets
\begin{equation}
f_{el}(\theta)=f_{C}(\theta)+ik\ \int_{0}^{\infty}db\ b\ J_{0}(qb)\ \exp
\left[  i\chi_{C}(b)\right]  \ \left\{  1-\exp\left[  i\chi_{N}(b)\right]
\right\}  \ . \label{simplif}%
\end{equation}

The advantage of using this formula is that the term $1-\exp\left[
i\chi _{N}(b)\right]  $ becomes zero for impact parameters larger
than the sum of the nuclear radii (grazing impact parameter).
Thus, the integral needs to be performed only within a small
range. In this formula, $\chi_{C}$ is given by eq. (\ref{intro4})
and $f_{C}(\theta)$ is given by eq. (\ref{fctheta}), with
\begin{equation}
\phi_{0}=-\eta C+\sum_{j=0}^{\infty}\left(  {\frac{\eta}{j+1}}-\arctan
{\frac{\eta}{j+1}}\right)  \ , \label{elast6}%
\end{equation}
where $C=0.5772156...$ is Euler's constant.

The elastic cross section can be expressed in terms of the
transverse momentum by using the relationships $d\Omega\simeq
d^{2}k_{\perp}/k^{2}$, and $k_{\perp}\simeq q=2k\sin\left(
\theta/2\right)  $, valid for high-energy collisions.

\section{Examples of momentum distributions}
\label{exam}We explore the consequences of the expressions
developed in sect.\,\ref{calc} by calculating momentum
distributions for selected cases. The longitudinal momentum
distributions, corresponding to a projection on the beam ($P_z$)
axis, turn out to be very close to those obtained in the simpler
black-disk approximation \cite{han95,han96,esb96}. For this
reason, a comparison with the numerous experimental data available
is hardly necessary, but some references are given. A general
discussion of longitudinal momentum distributions can be found in
\cite{han03}. For the distribution projected to an axis
perpendicular to the beam axis, the situation is different. We
find in all cases an interesting intermingling of momentum
components arising from stripping of the nucleon and from elastic
scattering  of the core fragment on the target. The latter
mechanism is diffractive for light targets and Coulomb-dominated
for heavy targets. There is very little useful experimental
evidence on the transverse momentum distributions, as only the
observation of coincident gamma rays can separate out the
differential cross sections to individual final levels. (Even for
halo nuclei, such as $^{11}$Be and $^{15}$C, 20--30\% of the
inclusive cross section goes to excited levels with different $l$
values.)

We have chosen to represent the momentum distributions graphically
in the following way. Projections onto a single Cartesian coordinate
are shown for one half axis only (the $P_z$ distribution is
symmetric in the eikonal approximation). For cases where the
magnetic quantum number $m$ differs from zero, we have weighted the
differential cross section with the multiplicity of 2, so that the
sum over all $m$ components gives the total cross section. The
spatial momentum distribution does not depend on the azimuthal
angle. It is therefore convenient to present it as a two-dimensional
function of the parallel and the transverse momentum with the
definition
\begin{equation}
\frac{d^2\sigma_{\mathrm{str}}}{dk_{\bot} dk_{z}}= 2 \pi k_{\bot}
\frac{d^3\sigma_{\mathrm{str}}} {d^{2}k_\bot dk_z} ,%
\label{sig2d}%
\end{equation}
which normalizes to the total cross section when the integration
is extended over the negative $k_z$ axis.
\begin{figure}[tbp]
\includegraphics[scale=0.8,angle=0]{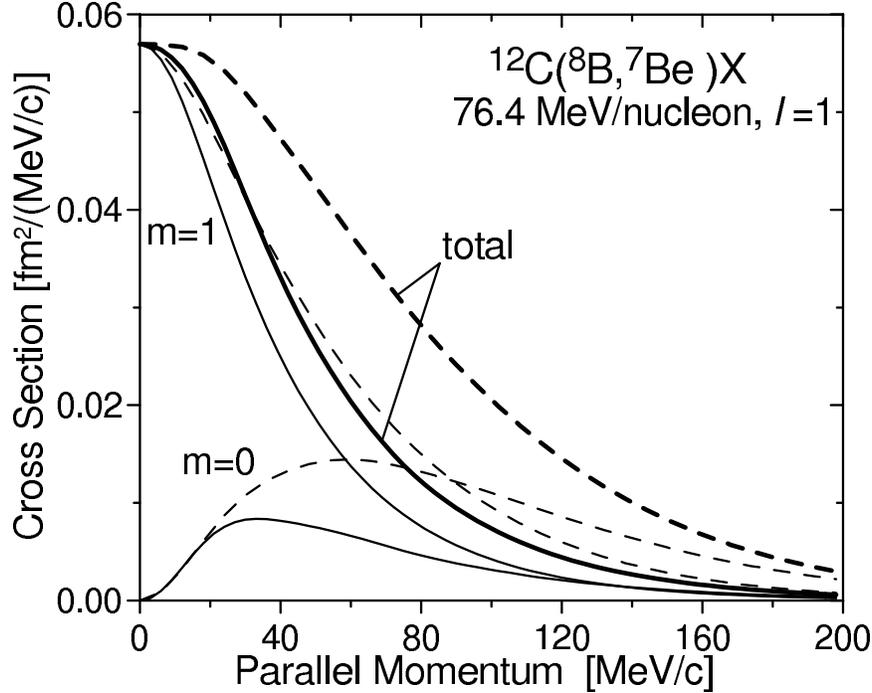}
\caption{ Longitudinal momentum distribution for the residue in
the $^{12}$C($^{8}$B,$^{7}$Be$_{gs}$)X reaction at 76 MeV/nucleon
as studied in the experiment of Enders {\textit et al.}
\cite{end03}. The proton binding energy for this proton halo state
is 0.137 MeV. The dashed curves are the cross sections calculated
in the transparent limit, re-scaled by a factor of 0.35. The
narrower full drawn curves are the results of the present work.
The thin curves show individual $m=\pm 1$ components with the sums
given as thick lines. The black disk calculation (not shown) gives
very similar results. Fig. 4 of ref.\,\cite{han03} compares a
similar
(black-disk) calculation with experimental data.}%
\label{b8z}
\end{figure}

\subsection{Longitudinal momentum distributions}
\label{long}The deuteron was the first halo system to be studied.
Serber \cite{Ser47} calculated the momentum spectrum of neutrons
from deuteron breakup on a light target in an approximation that
amounts to replacing the core S-matrix $S_{c}\left(  b_{c}\right)$
in eq. (\ref{sknock}) by unity. This leads to the expression%
\begin{equation}
\frac{d\sigma_{\mathrm{str}}}{d^{3}k_{c}}=\frac{1}{\left(  2\pi\right)  ^{3}%
}\frac{1}{2l+1}\ \sigma_{pT}\sum_{m}\left\vert \int d^{3}r\ e^{-i\mathbf{k}%
_{c}\mathbf{.r}}\psi_{lm}\left(  \mathbf{r}\right)  \right\vert ^{2}%
,\label{serber}%
\end{equation}
where $\sigma_{pT}$\ is the proton-target reaction cross section,
so that the stripping cross section (the term coined by Serber,
see his comment in \cite{ser94}) is that of a free proton, and the
momentum distribution is the Fourier transform of the bound-state
wave function. This is an excellent approximation for the
deuteron, and it served later to explain qualitatively why early
studies of stripping reactions of halo nuclei observed narrow
transverse and longitudinal momentum distributions
\cite{Ko88,Or92}.
\begin{figure}[tbp]
\includegraphics[scale=0.8,angle=0]{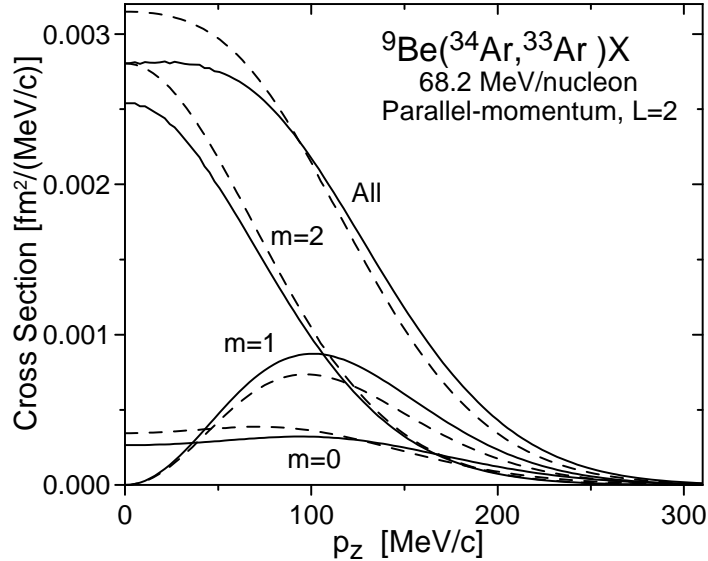}
\caption{ Longitudinal momentum distributions for the reaction $^{9}%
\mathrm{Be}(^{34}\mathrm{Ar},^{33}\mathrm{Ar}(\frac{3}{2}^+)) $ at
68.2 MeV/nucleon. This $l=2$ neutron-removal reaction leads to a
final $\frac{3}{2}^+$ level bound by 18.42 MeV. The solid curves
represent the exact calculations and the dashed curves the
sharp-cutoff
approximation.}%
\label{ar34z}
\end{figure}

However, the transparent-core limit is never a good approximation
for reactions of nuclei heavier than the deuteron. It
overestimates the stripping cross section by a factor 2 for a
pronounced halo system such as $^{11}$Be, and by up to a factor of
25 for deeply bound nucleons. The momentum distributions are also
modified in a significant way by the absorption given by the core
S-matrix $S_{c}\left( b_{c}\right)$ as shown by the examples of
the $l=0$ $^{11}$Be neutron halo and the $l=$1 $^{8}$B proton halo
shown in figs.\,\ref{be11z} and \ref{b8z}. The underlying reason
is that the knockout reaction of a heavy ion on an absorptive
target, such as $^{9}$Be, is surface dominated and samples only
the momentum components there and in the tail of the wave
function. The case of $^{11}$Be in fig.\,\ref{be11z} is good
demonstration of this: The-high momentum components (the
``shoulder") seen in the $S_{c}\left(  b_{c}\right) = 1$
approximation arise from the inner lobe in the wave function of
the second $s$ state and vanish almost completely in the knockout
reaction, which cannot ``see" the interior of the projectile.

In the case of the $l$=1 knockout on $^{8}$B, shown in fig.
\ref{b8z}, both $m$ components are much narrower than in the
transparent-core approximation, and the $m$=0 state is suppressed
relative to the $m$=$\pm$1 states. A simple geometrical argument
\cite{esb96} accounts for this: Projected onto the $(x,y)$ plane,
the two lobes of $m$=0 wave function are oriented along the $z$
axis and shielded by the core from interacting (alone) with the
target.

The sharp-cutoff model \cite{ann94,esb96,han96} has been a useful
tool for discussing parallel-momentum distributions. In this
approximation the S-matrices $S_{c}\left( b_{c}\right)$ and
$S_{n}\left( b_{n}\right)$ (see fig.\,\ref{coord}) are replaced by
step functions with radii chosen to approximate the free
core-target and nucleon-target cross sections by means of two
parameters, a target radius and a minimum-impact parameter. In the
following we have scaled the latter by a factor of typically 0.95
to get agreement with the, more accurate, total cross sections.
This adjustment has essentially no influence on the calculated
shapes. For the two halo cases discussed in figs.\,\ref{be11z} and
\ref{b8z} there is essentially no difference from the more exact
calculation. The same is true for the deeply bound $s$ state shown
later, in fig. \ref{ar34y0}.

The case of a deeply bound state with $l$=2 is shown in fig.
\ref{ar34z}. The reaction %
$^{9}\mathrm{Be}(^{34}\mathrm{Ar},^{33}\mathrm{Ar}(\frac{3}{2}^+))$ %
at 68.2 MeV/nucleon has a separation energy of 18.42 MeV and has
been studied experimentally by Gade {\em et al.} \cite{gad04}. The
solid curves are exact calculations and the dashed curves are
obtained with the sharp-cutoff approximation. In this case there
is a noticeable difference between the two approximations.

Finally, we draw attention to an interesting effect that clearly
is outside the scope of the present paper. The experiment on
$^{34}$Ar \cite{gad04} shows an excess of intensity at the
low-energy side of the $l$=0 and $l$=2 momentum distributions.
This cannot be accounted for in the eikonal approximation as
presented here, since it does not contain multistep time
dependence, does not conserve energy, and yields symmetric line
shapes. A similar low-momentum tail was observed in the $l=0$
momentum distributions from neutron knockout on the $^{11}$Be and
$^{15}$C halo states \cite{tos02}. In this case, it was
interpreted as arising from the diffractive reaction channel and
could be quantitatively accounted for in a coupled-channels
calculation. However, diffraction dissociation is a small
contribution to the cross section in the case of $^{34}$Ar and is
hardly the explanation here. We have no explanation for this
asymmetry, which has been seen also in other cases, and which
remains an interesting problem for future investigations.
\begin{figure}[tbp]
\begin{center}
\includegraphics[scale=0.65,angle=0]{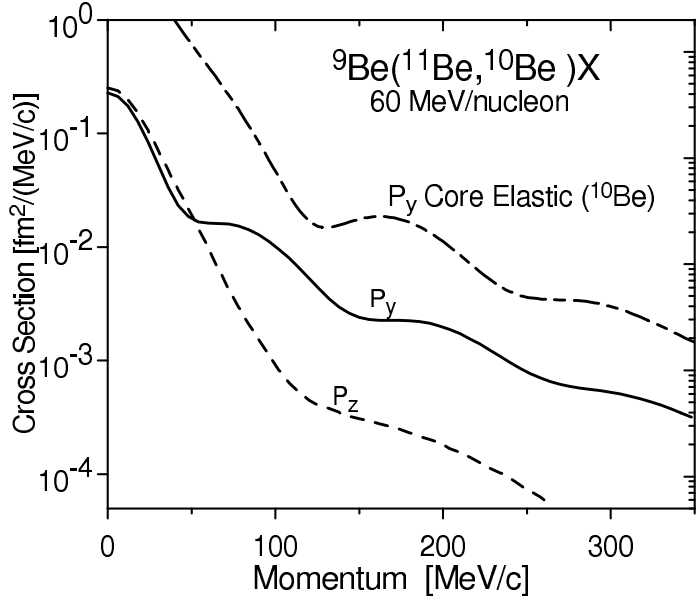}
\end{center}
\caption{ Transverse momentum distribution (full drawn) for the
reaction
$^{9}\mathrm{Be}\left(^{11}\mathrm{Be},^{10}\mathrm{Be_{gs}}\right)
$ at 60 MeV/nucleon. The corresponding longitudinal momentum
distribution (dashed) is shown for comparison. The dot-dashed
curve is the calculated elastic cross section for the core.}%
\label{be11y}%
\end{figure}

\subsection{Transverse momentum distributions}
Transverse momentum distributions depend more strongly on the
details of the nucleus-nucleus interaction than do the
longitudinal momentum distributions \cite{ber92}. The nuclear
size, the diffuseness of the nuclear matter distribution, and the
core-target Coulomb repulsion all contribute to the transverse
momentum distributions. At large impact parameters the Coulomb
force still has a strong influence, especially for heavy targets,
which for halo systems make Coulomb dissociation the dominant
breakup channel.

In figure \ref{be11y} we compare the longitudinal and transverse
momentum distributions for the reaction
$^{9}\mathrm{Be}\left(^{11}\mathrm{Be},^{10}\mathrm{Be}\right)$ at
60 MeV/nucleon. The transverse momentum distributions (solid
curve) has for small momenta the same shape as the longitudinal
one (dashed), but is lower in intensity by about 10\%. The missing
cross section shows up as a broad distribution with an oscillatory
pattern, which we interpret as elastic scattering of the core
fragment simultaneously with the stripping of the neutron. We
demonstrate this by calculating the core-target elastic cross
section, which shows a similar pattern with the minima
characteristic of Fraunhofer scattering. The minima are not in the
same place, however, because of the condition that the neutron be
absorbed in the stripping reaction. A similar broad component
appears in the transverse momentum distribution of the core
fragment in coupled-channels calculations (ref. \cite{tos02} and
J.A. Tostevin, personal communication).
\begin{figure}[tbp]
\begin{center}
\includegraphics[scale=0.8,angle=0]{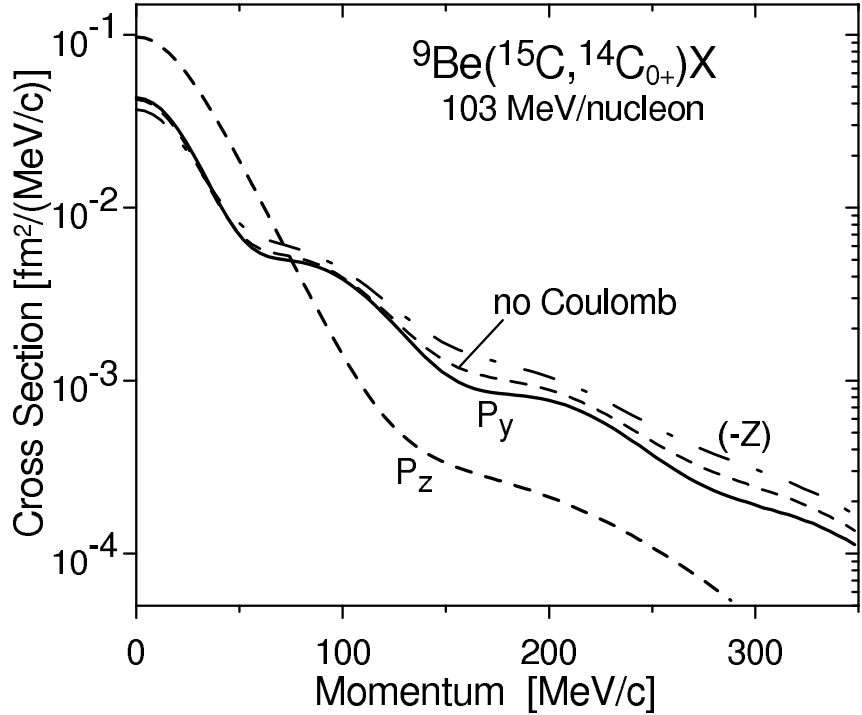}
\end{center}
\caption{ Transverse momentum distribution (full drawn) for the
reaction $^{9} \mathrm{Be}\left(
^{15}\mathrm{C},^{14}\mathrm{C_{gs}}\right)  $ at 103 MeV/nucleon.
The parallel-momentum distribution (dashed) has been studied
experimentally \cite{tos02,ter04}. The influence of the Coulomb
interaction is small, but visible with the log scale, for this
target-core combination. The curves labelled ``no Coulomb" and
``(-Z)" represent a calculation with the target charge set to
zero and to -4, respectively.}%
\label{c15y}%
\end{figure}

\begin{figure}[tbp]
\begin{center}
\includegraphics[scale=0.65,angle=0]{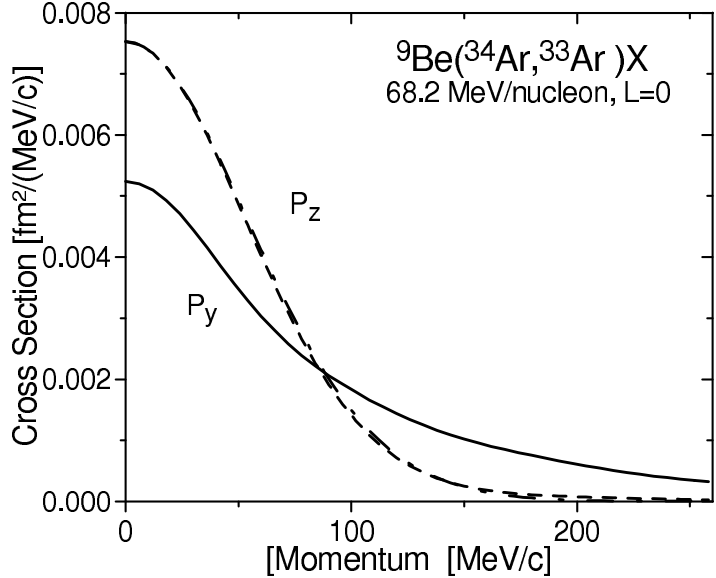}
\end{center}
\caption{ Transverse momentum distribution (full drawn) for the
$l$=0 knockout reaction
$^{9}\mathrm{Be}(^{34}\mathrm{Ar},^{33}\mathrm{Ar}(\frac{1}{2}^+)
$ at 68 MeV/nucleon and with a neutron separation energy of 17.06
MeV \cite{gad04}. The corresponding longitudinal momentum
distribution (dashed) is essentially indistinguishable from the
calculation in the black-disk approximation (dot-dashed).}%
\label{ar34y0}%
\end{figure}

\begin{figure}[tbp]
\begin{center}
\includegraphics[scale=0.65,angle=0]{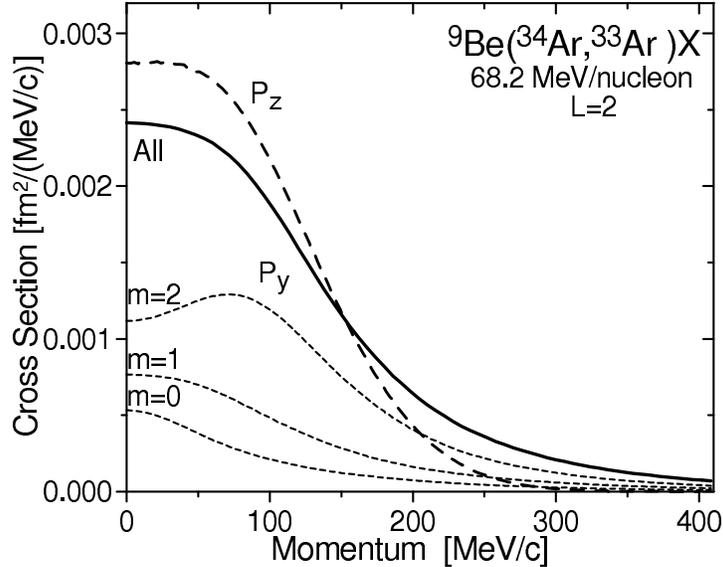}
\end{center}
\caption{ Transverse momentum distribution (full drawn) for the
$l$=2 knockout reaction
$^{9}\mathrm{Be}\left(^{34}\mathrm{Ar},^{33}\mathrm{Ar}\right) $
at 68 MeV/nucleon, corresponding to the $\frac{3}{2}^+$ excited
level in the core with an effective neutron separation energy of
18.43 MeV. The longitudinal momentum distribution (dashed) is
shown for comparison.}
\label{ar34y2}%
\end{figure}

For the case of $^{11}$Be, bound by 0.504 MeV, the large size of
the halo allows most absorption processes to occur without
simultaneous elastic scattering of the core. For more bound states
the two mechanisms become increasingly intermingled. Already in
the stripping of the halo neutron of $^{15}$C, bound by 1.218 MeV,
the tail from core scattering is much stronger, as seen in
fig.\,\ref{c15y}, and for the $l$=0 neutron knockout from
$^{34}$Ar, bound by 17.06 MeV, the two mechanisms are no longer
distinguishable, as seen in fig.\,\ref{ar34y0}.  The case of
knockout of a deeply bound $l$=2 neutron, also from $^{34}$Ar, is
shown in fig.\,\ref{ar34y2}, which again shows a transverse
momentum distribution that is broader than the longitudinal
distribution shown in fig.\,\ref{ar34y2}. Note the different
momentum dependence of the individual $m$ components, which can
easily be understood from geometrical properties of the spherical
harmonics \cite{esb96}.

The longitudinal momentum distributions are not affected by
elastic scattering of the core fragment. The basic reason for this
is that the forces acting on the core fragment are along the
z-direction and reverse sign at the origin of the $z$ axis, thus
leading to a null effect provided the angular deflection of the
core fragment is small. For the same reason, the longitudinal
momentum distributions are also relatively insensitive to Coulomb
effects.

We now examine how the Coulomb force affects the transverse
momentum distributions. In the case of a neutron halo, it pushes
the core away from the target along the transverse direction. The
resulting distributions for the case of a light target are here
illustrated by carbon ($Z$=6) on beryllium ($Z$=4), shown in
fig.\,\ref{c15y}. The effect is small and would hardly be
measurable. We note, however, that leaving out the Coulomb
interaction gives a broader momentum distribution, suggestive of
an interference effect. There is an interesting way to explore
this.

It is well known that the higher order Coulomb effects, e.g. to
second-order in perturbation theory, carry information on the
absolute sign of the charge of the particles. In atomic physics
this is often referred to as the Andersen-Barkas effect, which
reflects contributions of odd powers in the projectile charge on
stopping powers and ionization probabilities. In nuclear physics
this idea has been explored in the study of dynamical effects in
Coulomb dissociation \cite{EB02}. We have carried out calculations
reversing the sign of the target's charge but keeping all nuclear
interactions constant. Fig.\,\ref{c15y} shows that a hypothetical
core-target Coulomb attraction, as could be expected, gives an
even broader distribution. The reason for this is that the elastic
scattering at large transverse momenta is dominated by the
far-side scattering contribution, in which the nuclei pass by
close to each other. For a negatively-charged target both nuclear
and Coulomb forces interfere constructively, leading to larger
deflection angles.
\begin{figure}[tbp]
\begin{center}
\includegraphics[height=5in, width=3.2 in ]{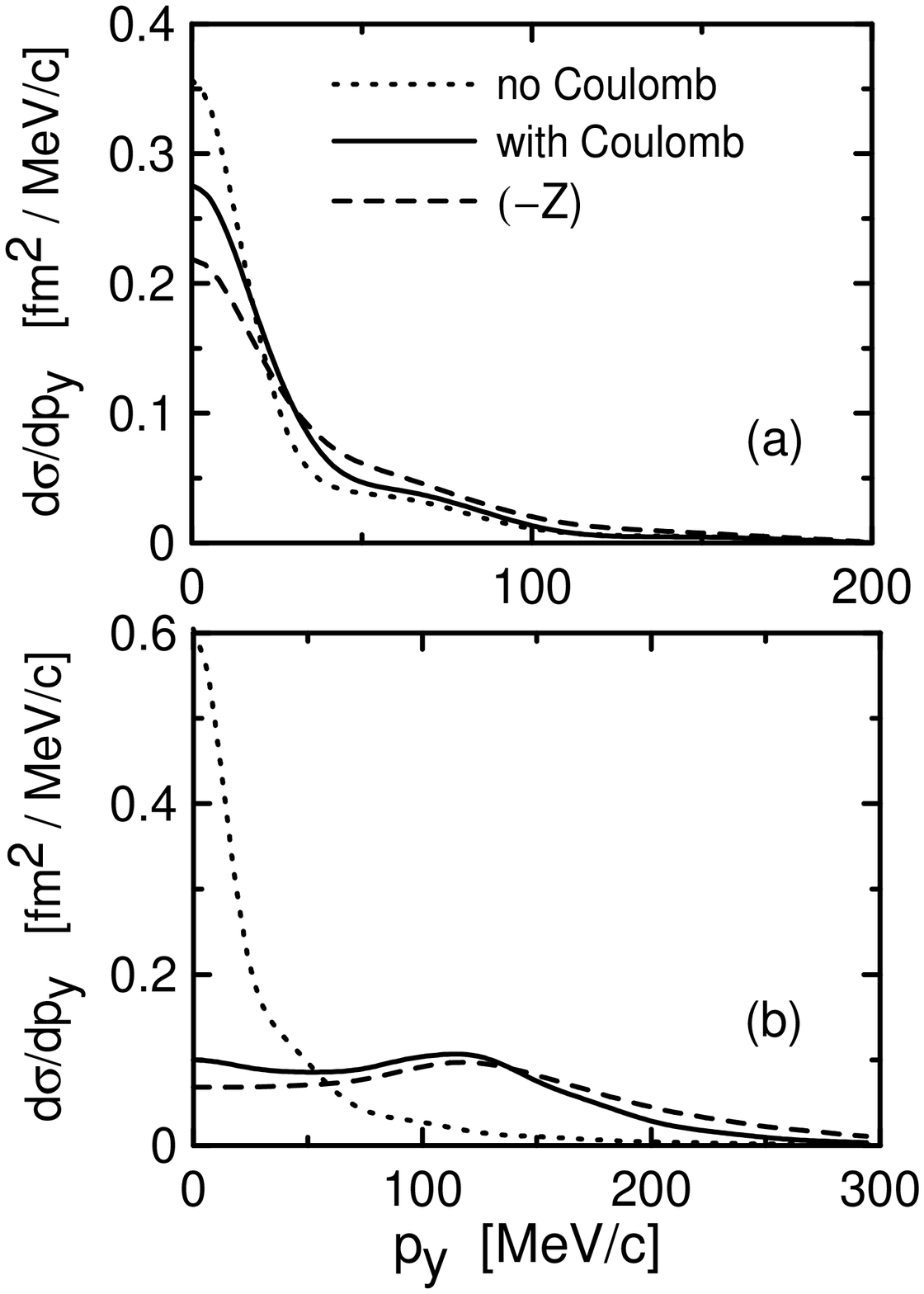}
\end{center}
\caption{Transverse momentum distributions of $^{10}\mathrm{Be}$
fragments for the reaction $\left(
^{11}\mathrm{Be},^{10}\mathrm{Be}\right)  $ at 80 MeV/nucleon. The
solid (dotted) curves include (do not include) the core-target
Coulomb interaction. The upper panel (a) is for Si targets, while
the lower panel (b) is for Pb targets. Changing the sign of the
charge of the
target yields the dashed curves.}%
\label{tsipb}%
\end{figure}
\begin{figure}[tbp]
\begin{center}
\includegraphics[scale=0.65,angle=0]{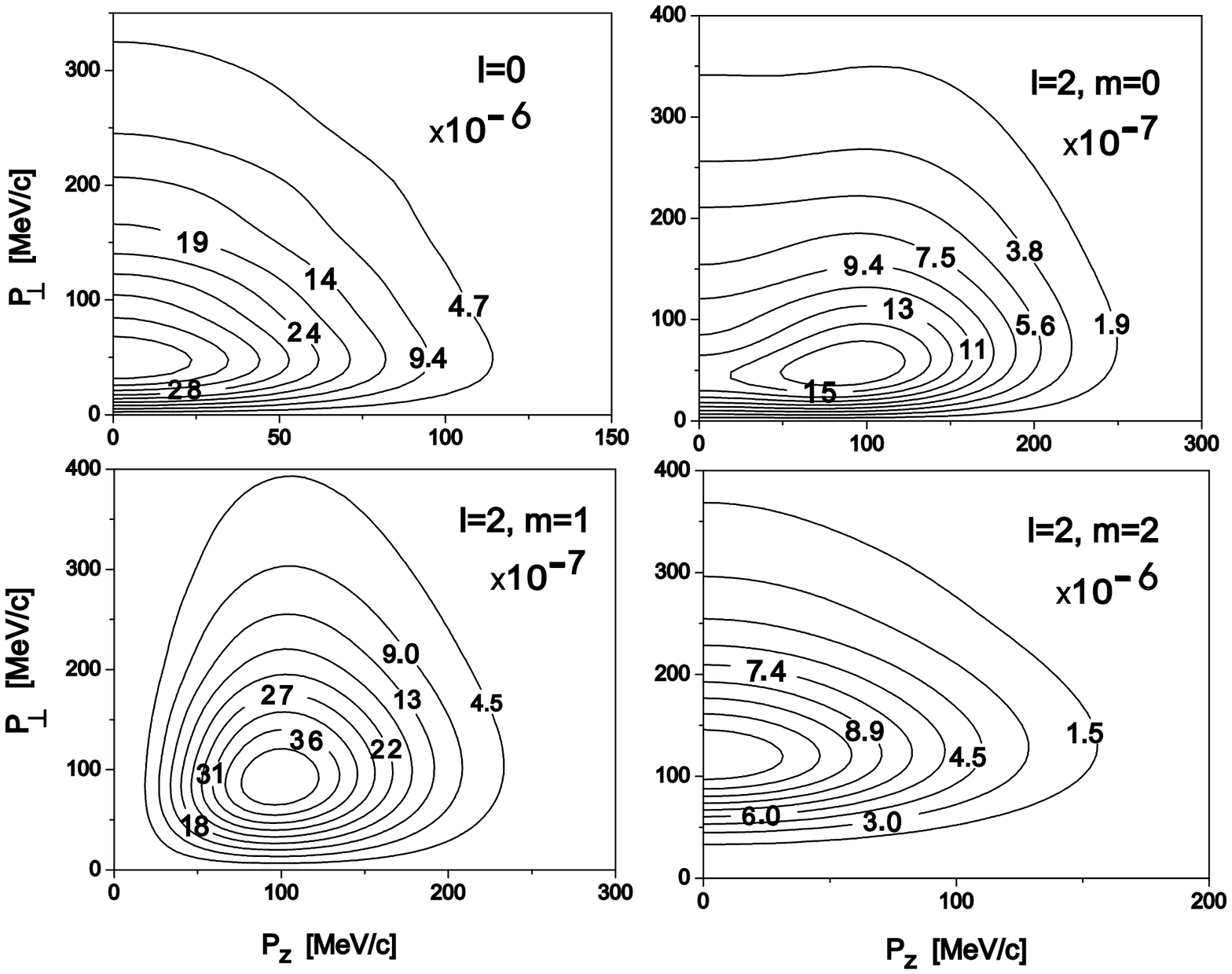}
\end{center}
\caption{ Contour plots for the $l$=0,2 knockout reactions
$^{9}\mathrm{Be}\left( ^{34}\mathrm{Ar},^{33}\mathrm{Ar}\right)X $
at 68.2 MeV/nucleon. The contour lines are equidistant. The
projections on one Cartesian coordinate axis were illustrated in
figs.\,\ref{ar34z}, \ref{ar34y0}, and \ref{ar34y2}. The absolute
values of $d^2 \sigma/ dk_z dk_\bot$ (in fm$^2$/(MeV/c)$^2$) are
given by the numbers in the contour plots, which are to be
multiplied by the factors shown in each panel.}
\label{contour}%
\end{figure}

The Coulomb effect becomes important for heavier targets. To
illustrate this, we compare in figure \ref{tsipb} the transverse
momentum distributions for the $(^{11}\mathrm{Be},^{10}\mathrm{Be})$
reaction on silicon (upper panel) and on lead (lower panel) targets.
One sees that the momentum distributions are somewhat distorted by
the Coulomb field for a silicon target, while for a lead target the
effect is huge and the outcome is completely dominated by Coulomb
repulsion. The transverse momentum transfer to the core fragments
after the stripping can be roughly estimated by the relation $\Delta
P_{c}=2Z_{c}Z_{T}e^{2}/bv$, where $b$ is the impact parameter and
$v$ is the projectile velocity. For the reaction on a lead target,
the minimum impact parameter is $b\simeq10$ fm, and with
$v\simeq0.4\ c$, we obtain $\Delta P_{c}\simeq220$ MeV/c. This
agrees well with the upper limit of the calculated distribution; the
broad peak (lower panel of figure \ref{tsipb}) is approximately one
half of that value.

Finally, we remind the reader that the present paper is an
analysis of the stripping reaction, also called inelastic breakup.
For experiments that only observe the core fragment, there is also
an incoherent contribution from elastic breakup, which is given by
the coherent sum of the contributions from diffraction
dissociation and Coulomb dissociation. The latter becomes dominant
for the reaction of $^{11}$Be \cite{ann94} on lead and will lead
to momentum distributions that, in principle, are different from
those shown in fig.\,\ref{tsipb}.

\section{Calculation of the double-differential stripping cross
 section and comparisons with experiment}
\label{ddif}%
\subsection{Alignment of the reaction residues}
Except for the special case in which the wave function factorizes
in Cartesian coordinates, the longitudinal and transverse momentum
components will be coupled. This is of primary importance for the
analysis of experiments in which the detection system limits the
acceptance of events to a certain volume in momentum space, a
problem that until now has been treated in simple approximations.
The selection in ($P_z$, $P_\bot$) space may also in special cases
provide an additional spectroscopic tool. In order to illustrate
the phenomenon, we plot in fig. \ref{contour} the double
differential cross section $d^2\sigma/dP_{z}dP_{\bot}$ for the
reaction of a deeply bound $l$=2 state calculated from eq.
(\ref{sig2d}). The momentum distributions for the $l=0$ and the
three $l=2$ states are shown separately. For the case $l$=2, the
distributions for the three $m$ states are peaked in different
regions of the $P_{z}$-$P_{\bot}$ plane. The maps of fig.
\ref{contour} are clearly necessary input to an accurate
calculation of the experimental acceptance.

\begin{figure}[tbp]
\begin{center}
\includegraphics[scale=1.0,angle=0]{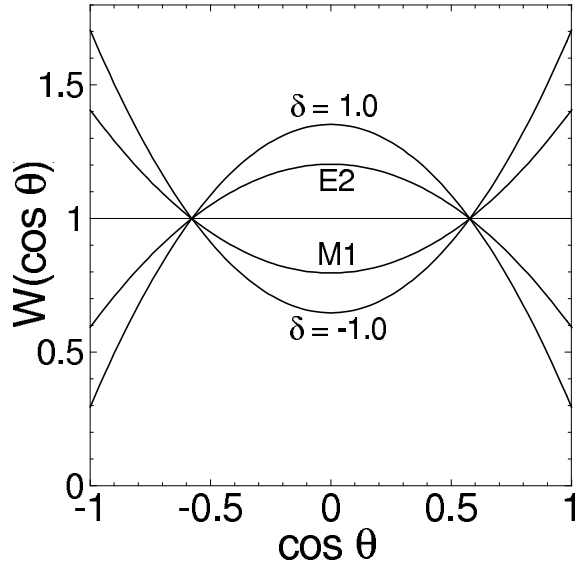}
\end{center}
\caption{ Gamma-ray angular distributions for a $\frac{3}{2}^+$ to
$\frac{1}{2}^+$ transition with alignment parameters corresponding
to ``set 2" (in the text) obtained by a central cut on the
parallel momentum. Calculations are shown for the multipolarities
E2, M1,
and for mixed transitions with E2/M1 amplitudes of $\pm$1.0.}%
\label{o22}%
\end{figure}

We have already pointed out that the reactions favor the formation
of residues in states with the maximum absolute value of the
magnetic quantum number $m=\pm l$ (with the quantization axis
taken to be the beam direction).  This (possible) alignment effect
implies a (possible) anisotropic emission (in the center-of-mass
system) of gamma rays emitted after the reaction. This is a tool
for identifying multipolarities of the gamma transitions and spin
sequences in the product nucleus. The fact that for different
values of $m$, the main contributing cross sections are located in
different areas of the ($P_z$, $P_\bot$) map suggests that the
alignment effect can be enhanced by making cuts in the momentum
components. A theoretical example of such an application has been
given in fig. 12 of ref. \cite{han03}, which shows the calculated
angular distributions for two different spin sequences. For the
example given in fig.\,\ref{o22}, the $m$=2 fraction in the
reaction residues is (``set 1") 58\% as compared with 40\% for a
population with statistical weights. Selecting reactions with
$P_z$ values between -50 and 50 MeV/c (``set 2") increases the
$m$=2 fraction to 85\% and reduces the count rate to one half,
corresponding to a net gain in sensitivity. Limiting in addition
the values of $P_\bot$ to values between 85 and 165 MeV/c gives
only a marginal improvement: The $m$=2 fraction increases to 88\%
but at the cost of reducing the intensity by another factor of one
half.

As an example of a possible application of the alignment effect,
we show in fig.\,\ref{o22} angular distributions of gamma rays in
the center-of-mass system calculated from the expressions given by
Yamazaki \cite{yam67}. The example is for an assumed
$\frac{3}{2}^+$ to $\frac{1}{2}^+$ cascade as in  $^{33}$Ar and
for various assumed multipoles and mixing ratios. The alignment
parameters correspond to ``set 2", which gives results that are
already approaching pure $m$=$\pm$2 alignment. Without the cut on
$P_z$ the anisotropy would be only half as big.

\subsection{Comparisons with experiment}
The present work deals with an extension of a theory
\cite{tos99,tos01} that has been tested in numerous experiments. As
indicated in subsect.\,\ref{long} the longitudinal momentum
distributions are well in hand, both experimentally and
theoretically. There is little reason to doubt that the same is the
case for the transverse-momentum distributions calculated here.
However, there are few good data sets to compare with. Essentially
all measured transverse distributions that we are aware of are
superpositions of contributions from several $l$ values (see the
large set of inclusive momentum spectra reported by Sauvan {\em et
al.} \cite{sau04}). For these, coincidence measurements with gamma
rays would be required to provide detailed test of our calculations.
However, the proton knockout on $^{8}$B is an exception to this.

\begin{figure}[tbp]
\begin{center}
\includegraphics[scale=0.7,angle=0]{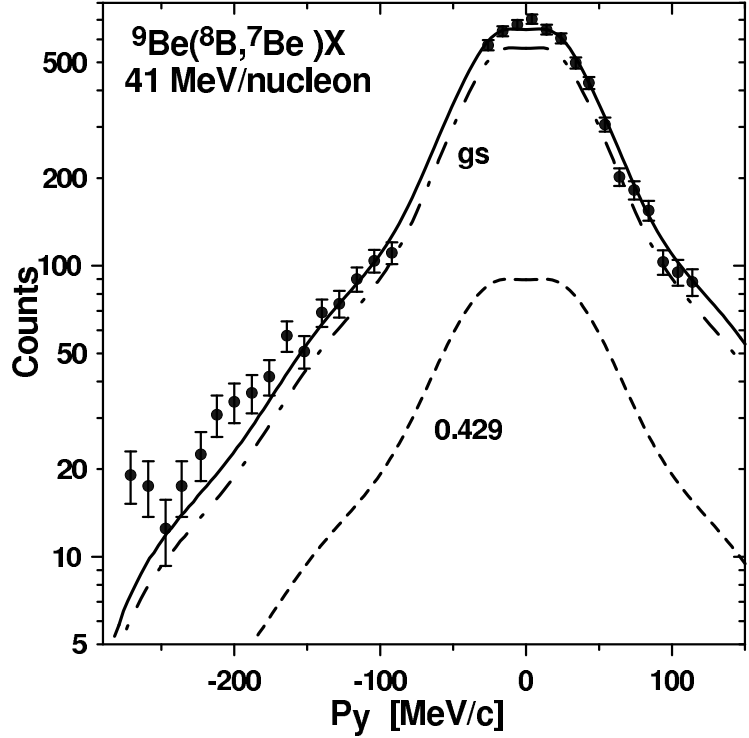}
\end{center}
\caption{Inclusive transverse-momentum distribution for the
residue in the $^{9}$Be($^{8}$B,$^{7}$Be$_{gs}$)X reaction
measured at 41 MeV/nucleon \cite{kel96}. The theoretical
calculation (full drawn) is based on the same parameter set as
fig.\,\ref{b8z}, and the only adjustable parameter is the
(arbitrary) scale. The binding energy for this proton halo state
is 0.137 MeV. The dot-dashed and dashed curves are the individual
contributions of the $^{7}$Be ground state and first-excited
state, respectively. The angular resolution in the experiment
broadens the data by approximately 4\%. This has not been included
in the theoretical curves.}
\label{b8y}%
\end{figure}

For $^{8}$B, the inclusive transverse-momentum distribution in the
knockout of the halo proton has been measured in the experiment by
Kelley {\textit et al.} \cite{kel96}. In spite of the absence of
gamma-ray coincidence data, this is a favorable case because the
ground-state cross section dominates and because the approximately
15\% branch to the excited level also has $l$=1 and has an almost
identical shape.  Fig.\,\ref{b8y} shows that our calculation is in
excellent agreement with the data, which it reproduces over two
orders of magnitude in cross section. There are several
measurements of the longitudinal-momentum distribution for $^{8}$B
proton knockout; fig. 4 of ref.\, \cite{han03} presents a
comparison of high-energy data with a black-disk calculation,
essentially indistinguishable from fig.\,\ref{b8z} of the present
work.

Our second comparison with experimental data illustrates how the
calculations in the present work may be applied toward clarifying
a much more complicated issue. The reactions and structure of the
two-neutron halo nucleus $^{11}$Li have attracted much interest.
It is a Borromean system in the sense that although the three-body
system, consisting of $^{9}$Li and two neutrons, forms a bound
state, none of the possible two-body subsystems have bound states.
Hence the stability of $^{11}$Li is brought about by the interplay
of the core-neutron and the neutron-neutron interactions, which
must lead to a strongly correlated wave function with the two
neutrons spatially close together. Barranco {\em et al.}
\cite{bar93} found that attempts to understand $^{11}$Li via
breakup reactions are made difficult by important final-state
interactions, so that the primary mechanism is removal of a single
neutron followed by the decay in flight of the unstable $^{10}$Li.
The slow neutron emerging from this decay carries no direct
information on the $^{11}$Li structure; its energy spectrum
reflects properties of both the initial and final state.

\begin{figure}[tbp]
\begin{center}
\includegraphics[scale=0.6,angle=0]{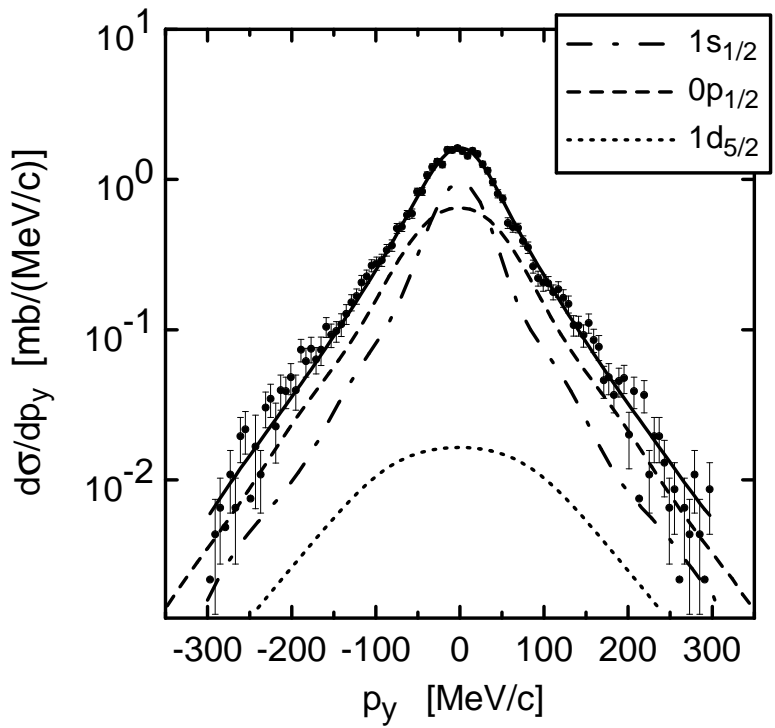}
\end{center}
\caption{Inclusive transverse-momentum distribution for the
residue in the inclusive $^{12}$C($^{11}$Li,$^{10}$Li)X reaction
measured at 287 MeV/nucleon \cite{sim99}. The theoretical result
(full drawn curve) is the adjusted sum of three angular-momentum
components as described in the text.}
\label{p10li}%
\end{figure}

Simon {\em et al.} \cite{sim99} showed that this problem can be
circumvented by reconstructing the momentum vector of the
$^{10}$Li intermediate from the observed momenta of $^{9}$Li and a
neutron following a reaction on a carbon target. In an experiment
at 287 MeV/nucleon they obtained the best resolution for the
projected transverse-momentum component and arrived at the
spectrum shown in fig.\,\ref{p10li}. The momentum resolution (full
width at half maximum) was approximately 55 MeV/c \cite{sim04}.
Their analysis was based on a simplified model using analytical
approximations to the momentum distributions and found the
spectrum to require an $(1s_{1/2})^2$ component of (45$\pm$10)\%,
the rest being $(0p_{1/2})^2$. They obtained further confirmation
of this interpretation by measuring the angular distribution of
the neutron relative to a quantization axis taken along the recoil
direction of the $^{10}$Li composite. This distribution showed the
forward-backward asymmetry characteristic of interference between
final states of opposite parity.

For the calculation here we approximate the $^{11}$Li ground state
by an inert $^{9}$Li core coupled to a neutron pair in a mixture
of $(1s_{1/2})^2$, $(0p_{1/2})^2$, and $(0d_{5/2})^2$ states. We
assume a two-body model, thus neglecting interference effects, and
adjust the single-particle wave functions to reproduce the
effective neutron separation energies. The two-neutron separation
energy is 0.3 MeV and to this comes the average excitation
energies of the three states in $^{10}$Li. From the systematics in
fig. 6 of \cite{han01} augmented with data for the $0d_{5/2}$
neutron states in neighboring nuclei we estimate $^{10}$Li average
excitation energies to 0.2, 0.5, and 1.5 MeV for the three
single-particle states, respectively. In the spirit of the sudden
approximation, we take these to be the center of gravity for the
multiplets formed by the coupling to the $^{9}$Li spin and arrive
at effective neutron-separation energies of 0.5, 0.8, and 1.8 MeV,
still in the same order, which determine the relative
core-particle wave functions of the system $^{10}$Li+n. The sudden
approximation suggests that the same wave functions should be used
for calculating the $S$ matrices (profile functions) for $^{10}$Li
treating it as a neutron+$^{9}$Li composite as described in
ref.\,\cite{tos99}. The three spectroscopic factors are then
obtained from an adjustment to the differential cross section of
fig.\,\ref{p10li}.

The calculated transverse-momentum distributions for the assumed
three components were folded with the experimental resolution. The
sum of the resulting distributions, weighted with the unknown
spectroscopic factors, were adjusted in a $\chi^2$ analysis to the
experimental result as shown in fig.\,\ref{p10li}. Per degree of
freedom, the goodness of the fit is $\chi^{2}/\nu=1.38$ for
absolute spectroscopic factors of 0.98(6) ($1s_{1/2}$), 1.91(16)
($0p_{1/2}$), and 0.12(10) ($0d_{5/2}$). In view of the many
approximations made in this analysis, it is probably satisfactory
that the the sum of the spectroscopic factors is 3.0, where the
sum-rule value is 2. Re-normalized to the sum, the relative
contributions are 33(2)\%, 64(5)\%, and 4(3)\% in approximate
agreement with the analysis of Simon {\em et al.} \cite{sim99}
cited above, who obtained 45\% for the $1s_{1/2}$ state. The
errors cited do not include systematic contributions arising from
the theory or contained in the experimental data. It is
interesting to compare with the neighboring nucleus $^{12}$Be
\cite{nav00}, also with 8 neutrons, for which the $0d_{5/2}$
contribution to the one-neutron knockout reaction appears to be
much larger, of the order of 50\%.

\section{Concluding remarks}
The present work has extended the theory of one-nucleon stripping
reactions (inelastic breakup) to cover two-dimensional momentum
distributions of the reaction residue with the use of realistic
profile functions for the core-target and nucleon-target
interactions. The parallel-momentum distributions, projections on
the beam direction $P_z$, are not appreciably different from those
obtained in the ``black disk" approximation used in earlier work.
On the other hand, the projections on the transverse direction,
here referred to as $P_y$, are very different from the projections
on $P_z$. For halo systems they show a weak component, which
represents an additional mechanism in which the residue has
scattered elastically on the target. For more strongly bound
systems the two processes, stripping and elastic scattering,
become inseparable, and the distributions on $P_y$ are broader
than those on $P_z$.

The doubly-differential distribution of the cross section on the
parallel- and transverse-momentum variables ($P_z$,$P_\bot$) has
for a given angular momentum $l\geq 1$ a characteristic behavior
for the different components of the magnetic quantum number $m$.
This will be of importance for evaluating acceptance corrections
in experiments. The $m$ dependence of the cross sections leads to
alignment of the residues and consequently to anisotropic emission
of subsequent gamma rays. This effect can be exploited for
identifying spin sequences and gamma-ray multipolarities as
illustrated in the example in fig.\,\ref{o22}. A measurement of
the transverse momentum distribution in stripping of $^{8}$B on a
light target is in excellent agreement with our calculation, which
reproduces the experimental result accurately over two orders of
magnitude in cross section. We have also attempted a more tenuous
application of the theory to the complex case of neutron knockout
on the two-neutron halo nucleus $^{11}$Li.

\begin{acknowledgments}
We have benefited from comments by Henning Esbensen, Kai Hencken
and Jeffrey A. Tostevin. This work was supported by the National
Science Foundation under Grants No. PHY-0110253 and PHY-0070911.
\end{acknowledgments}

\section{Appendix: Gaussian expansion method for stripping reactions}
Using the explicit form of the spherical harmonics%
\begin{eqnarray}
Y_{lm}\left(  \widehat{\mathbf{r}}\right)   &  =& \left(  -1\right)  ^{m}%
\sqrt{\frac{2l+1}{4\pi}}\sqrt{\frac{\left(  l-m\right)  !}{\left(  l+m\right)
!}}\ P_{lm}\left(  \cos\theta\right)  \ e^{im\phi}\ \nonumber\\
&  =&C_{lm}P_{lm}\left(  \cos\theta\right)  \ e^{im\phi} \label{Ylm}%
\end{eqnarray}
and
\begin{equation}
\mathbf{k}_{c}\cdot\mathbf{r}=k_{\bot}r\sin\theta\cos\left(  \phi_{k}%
-\phi\right)  +k_{z}r\cos\theta\ .
\end{equation}
Part of the integral in (\ref{sknock}) is
\begin{eqnarray}
\mathcal{F}_{lm}\left(  k_{\bot},k_z,b_{n}\right)   & =&\int
d^{3}r\ e^{-i\mathbf{k}_{c}\mathbf{.r}}S_{c}\left( b_{c}\right)
\psi
_{lm}\left(  \mathbf{r}\right) \nonumber\\
&  =&C_{lm}\int dr\ r^{2}\ \sin\theta\ d\theta\ d\phi\nonumber\\
&  \times&\exp\left\{  -i\left[  k_{\bot}r\sin\theta\cos\left(  \phi_{k}%
-\phi\right)  +k_{z}r\cos\theta\right]  \right\} \nonumber\\
&  \times& S_{c}\left(
\sqrt{r^{2}\sin^{2}\theta+b_{n}^{2}-2r\sin\theta
\ b_{n}\cos\left(  \phi-\phi_{n}\right)  }\right) \nonumber\\
&  \times& R_{l}\left(  r\right)  \ P_{lm}\left(
\cos\theta\right)
\ e^{im\phi}\ . \label{merd1}%
\end{eqnarray}
To simplify the calculations we can express $S_{c}\left(  b_{c}\right)  $ as
an expansion in terms of integrable functions. The S-matrices can be well
described by the expansion%
\begin{equation}
S_{c}\left(  b_{c}\right)  =\sum_{j}^{N}\alpha_{j}\ \exp\left[  -b_{c}%
^{2}/\beta_{j}^{2}\right]  \ ,\ \ \ \ \ \ \text{with}\ \ \ \beta_{j}%
=\frac{R_{L}}{j}\ . \label{scbc}%
\end{equation}
Good fits for realistic S-matrices were obtained with $N=20$, i.e. with 20
complex coefficients $\alpha_{j}$\ and $R_{L}=10-20$ fm, depending on the size
of the system. Since the real part of the S-matrices has the usual behavior of
$S_{c}\left(  b_{c}\right)  \sim0$ for $b_{c}\ll R$, and $S_{c}\left(
b_{c}\right)  \sim1$ for $b_{c}\gg R$, where $R$ is a generic nuclear size,
one of the coefficients of the expansion in eq. (\ref{scbc}) is $\alpha_{j}%
=1$, and $\beta_{j}=\infty$, which we take as the $j=0$ term in the expansion.

We now use the sum (\ref{scbc}) and the equation%
\begin{equation}
\exp\left(  -iz\cos\phi\right)  =\sum_{p=-\infty}^{\infty}i^{p}J_{p}\left(
z\right)  \ e^{ip\phi}\ , \label{Bexp}%
\end{equation}
valid for any complex $z$, in equation (\ref{merd1}).

The integration in eq. (\ref{merd1}) will then involve functions of the form%
\begin{eqnarray}
\mathcal{F}_{lm,\ j}\left(  k_{\bot},k_z,b_{n}\right)   &
=&C_{lm}\ \alpha_{j}\int d\rho\ \rho\ dz\ d\phi\nonumber\\
&  \times& R_{l}\left(  r\right)  \ P_{lm}\left(
\cos\theta\right) \ \exp\left[  -\left(
\mathbf{\mbox{\boldmath$\rho$}-b}_{n}\right)
^{2}/\beta_{j}^{2}\right] \nonumber\\
&  \times& e^{im\phi}\ \exp\left\{  -ik_{\bot}\rho\cos\left(  \phi_{k}%
-\phi\right)  \right\}  \ \exp\left[  -ik_zz\right]  \ ,
\end{eqnarray}
where $\mathbf{r}\equiv\left(  \mathbf{\mbox{\boldmath$\rho$}},z\right)  $,
$\cos\theta=z/r=z/\sqrt{\rho^{2}+z^{2}}$. Then%
\begin{eqnarray}
\mathcal{F}_{lm,\ j}\left(  k_{\bot},k_z,b_{n}\right)   &
=&C_{lm}\ \alpha_{j}\ \exp\left[  -b_{n}^{2}/\beta_{j}^{2}\right]
\int
d\rho\ \rho\ \exp\left[  -\rho^{2}/\beta_{j}^{2}\right]  \ \nonumber\\
&  \times&\int dz\ \exp\left[  -ik_zz\right]  \ R_{l}\left(
r\right)
\ P_{lm}\left(  \cos\theta\right) \nonumber\\
&  \times&\int d\phi\ e^{im\phi}\ \exp\left\{
-ik_{\bot}\rho\cos\left(
\phi_{k}-\phi\right)  \right\}  \ \exp\left[  \frac{2\rho b_{n}}{\beta_{j}%
^{2}}\cos\left(  \phi-\phi_{n}\right)  \right]  \ .
\end{eqnarray}

Using the expansion (\ref{Bexp}), we can write%
\begin{eqnarray}
& & \exp\left\{  -ik_{\bot}\rho\cos\left(  \phi_{k}-\phi\right)
\right\} \ \exp\left[  \frac{2\rho b_{n}}{\beta_{j}^{2}}\cos\left(
\phi-\phi
_{n}\right)  \right] \nonumber\\
&  =&\sum_{p,\ p^{\prime}}i^{p+p^{\prime}}J_{p}\left(
k_{\bot}\rho\right) \ J_{p^{\prime}}\left(  i\frac{2\rho
b_{n}}{\beta_{j}^{2}}\right) \ \exp\left[  ip\left(
\phi_{k}-\phi\right)  \right]  \ \exp\left[ ip^{\prime}\left(
\phi-\phi_{n}\right)  \right]  \ .
\end{eqnarray}
Since%
\[
\int d\phi\ e^{im\phi}e^{-ip\phi}e^{ip^{\prime}\phi}=2\pi\delta
_{m-p,\ p^{\prime}\ ,}%
\]
then%
\begin{eqnarray}
& & \int d\phi\ e^{im\phi}\ \exp\left\{  -ik_{\bot}\rho\cos\left(
\phi
_{k}-\phi\right)  \right\}  \ \exp\left[  \frac{2\rho b_{n}}{\beta_{j}^{2}%
}\cos\left(  \phi-\phi_{n}\right)  \right] \\
&  =&2\pi i^{m}\exp\left[  -im\phi_{n}\right]  \sum_{p}J_{p}\left(
k_{\bot }\rho\right)  J_{m-p}\left(  i\frac{2\rho
b_{n}}{\beta_{j}^{2}}\right) \exp\left[  ip\left(
\phi_{k}+\phi_{n}\right)  \right]  \ \ .
\end{eqnarray}

Thus%
\begin{eqnarray}
\mathcal{F}_{lm,\ j}\left(  k_{\bot},k_z,b_{n}\right)   & =&2\pi \
C_{lm}\ i^{m}\ \alpha_{j}\ \exp\left[  -im\phi_{n}\right] \
\exp\left[
-b_{n}^{2}/\beta_{j}^{2}\right] \nonumber\\
&  \times&\sum_{p}\exp\left[  ip\left(  \phi_{k}+\phi_{n}\right)
\right]
\nonumber\\
&  \times&\int_{0}^{\infty}d\rho\ \rho\ J_{p}\left(
k_{\bot}\rho\right) \ \exp\left[  -\rho^{2}/\beta_{j}^{2}\right]
\ J_{m-p}\left(  i\frac{2\rho
b_{n}}{\beta_{j}^{2}}\right) \nonumber\\
&  \times&\int_{-\infty}^{\infty}dz\ \exp\left[ -ik_zz\right]
\ R_{l}\left(  r\right)  \ P_{lm}\left(  \cos\theta\right)  \ . \label{fmd}%
\end{eqnarray}

Upon squaring eq. (\ref{fmd}), inserting in eq. (\ref{sknock}), and
integrating over $\phi_{k}$ and $\phi_{n}$, we can use $\int d\phi\ e^{ip\phi
}e^{-ip^{\prime}\phi}=2\pi\delta_{p,\ p^{\prime}}$ and we get%
\begin{equation}
\frac{d\sigma_{\mathrm{str}}}{k_{\bot}dk_{\bot}dk_z}=\ \frac{2\pi}%
{2l+1}\int_{0}^{\infty}db_{n}\ b_{n}\ \left[  1-\left\vert
S_{n}\left( b_{n}\right)  \right\vert ^{2}\right]  \ \sum_{m,\
p}C_{lm}^{2}\ \left\vert \mathcal{A}_{lmp}\left(
k_{\bot},k_z,b_{n}\right)  \right\vert ^{2},
\label{fknock}%
\end{equation}
where%
\begin{eqnarray}
\mathcal{A}_{lmp}\left(  k_{\bot},k_z,b_{n}\right)   &  =&\sum_{j=0}%
^{N}\alpha_{j}\ \exp\left[  -b_{n}^{2}/\beta_{j}^{2}\right] \nonumber\\
&  \times&\int_{0}^{\infty}d\rho\ \rho\ J_{p}\left(
k_{\bot}\rho\right) \exp\left[  -\rho^{2}/\beta_{j}^{2}\right]  \
J_{m-p}\left(  i\frac{2\rho
b_{n}}{\beta_{j}^{2}}\right) \nonumber\\
&  \times&\int_{-\infty}^{\infty}dz\ \exp\left[ -ik_zz\right]
\ R_{l}\left(  r\right)  \ P_{lm}\left(  \cos\theta\right)  \ . \label{almf}%
\end{eqnarray}

Since%
\[
I_{\alpha}\left(  x\right)  =i^{-\alpha}J_{\alpha}\left(  ix\right)  ,
\]
where $I_{\alpha}\left(  x\right)  $\ is the modified Bessel function, one
gets%
\begin{eqnarray}
\mathcal{A}_{lmp}\left(  k_{\bot},k_z,b_{n}\right)   &  =&\sum_{j}%
\alpha_{j}\ \exp\left[  -b_{n}^{2}/\beta_{j}^{2}\right] \nonumber\\
&  \times&\int_{0}^{\infty}d\rho\ \rho\ J_{p}\left(
k_{\bot}\rho\right) \exp\left[  -\rho^{2}/\beta_{j}^{2}\right]  \
I_{m-p}\left(  \frac{2\rho
b_{n}}{\beta_{j}^{2}}\right) \nonumber\\
&  \times&\int_{-\infty}^{\infty}dz\ \exp\left[ -ik_zz\right]
\ R_{l}\left(  r\right)  \ P_{lm}\left(  \cos\theta\right)  \ , \label{almf2}%
\end{eqnarray}
where an irrelevant phase $i^{m-p}$ was dropped off, as only the absolute
value of $\mathcal{A}_{lmp}$ enters eq. (\ref{fknock}).

The first term of the equation (\ref{almf2}), with
$\beta_{j}=\infty$ and $\alpha_{j}=1$ can be calculated using
$I_{\alpha}\left(  0\right) =\delta_{\alpha}$.

Using the integral%
\begin{equation}
\int_{0}^{\infty}dk_{\perp}\ k_{\perp}\ J_{p}\left(  k_{\bot}\rho\right)
\ J_{p}\left(  k_{\bot}\rho^{\prime}\right)  =\frac{1}{\rho}\ \delta\left(
\rho-\rho^{\prime}\right)  \ ,
\end{equation}
in eq. (\ref{fknock}) one gets for the \textit{longitudinal momentum
distribution}%
\begin{equation}
\frac{d\sigma_{\mathrm{str}}}{dk_z}=\
\frac{2\pi}{2l+1}\int_{0}^{\infty }db_{n}\ b_{n}\ \left[
1-\left\vert S_{n}\left(  b_{n}\right)  \right\vert ^{2}\right]  \
\sum_{m,\ p}C_{lm}^{2}\ \int_{0}^{\infty}d\rho\ \rho \ \left\vert
\mathcal{B}_{lmp}\left(  k_z,b_{n},\rho\right)  \right\vert
^{2}, \label{siglong}%
\end{equation}
where%
\begin{eqnarray}
\mathcal{B}_{lmp}\left(  k_z,b_{n},\rho\right)   &
=&\sum_{j}\alpha _{j}\ \exp\left[  -b_{n}^{2}/\beta_{j}^{2}\right]
\ \exp\left[  -\rho
^{2}/\beta_{j}^{2}\right]  \ \\
&  \times& I_{m-p}\left(  \frac{2\rho b_{n}}{\beta_{j}^{2}}\right)
\int_{-\infty}^{\infty}dz\ \exp\left[  -ik_zz\right]  \
R_{l}\left( r\right)  \ P_{lm}\left(  \cos\theta\right)  \
.\nonumber
\end{eqnarray}

Using the integral of eq. (\ref{DiracZ}) in eq. (\ref{fknock}) one gets for
the \textit{transverse momentum distribution}%
\begin{equation}
\frac{d\sigma_{\mathrm{str}}}{d^{2}k_{\bot}}=\ \frac{2\pi}{2l+1}\ \int
_{0}^{\infty}db_{n}\ b_{n}\ \left[  1-\left\vert S_{n}\left(  b_{n}\right)
\right\vert ^{2}\right]  \ \sum_{m,\ p}C_{lm}^{2}\ \int_{-\infty}^{\infty
}dz\ \left\vert \mathcal{D}_{lmp}\left(  k_{\bot},b_{n},z\right)  \right\vert
^{2}, \label{sigtrans}%
\end{equation}
where
\begin{eqnarray}
\mathcal{D}_{lmp}\left(  k_{\bot},b_{n},z\right)   &  =&\sum_{j=0}^{N}%
\alpha_{j}\ \exp\left[  -b_{n}^{2}/\beta_{j}^{2}\right] \\
&  \times&\int_{0}^{\infty}d\rho\ \rho\ J_{p}\left(
k_{\bot}\rho\right) \exp\left[  -\rho^{2}/\beta_{j}^{2}\right]  \
I_{m-p}\left(  \frac{2\rho b_{n}}{\beta_{j}^{2}}\right)
R_{l}\left(  r\right)  \ P_{lm}\left( \cos\theta\right)  \
.\nonumber
\end{eqnarray}
The formulas above are also useful to check the quality of the
Gaussian fit, eq. (\ref{scbc}), to obtain the momentum
distributions. One can compare the direct numerical integrations
using eq.\,(\ref{siglong}) with
\begin{eqnarray}
\frac{d\sigma_{\mathrm{str}}}{dk_z}  &  =&\ \frac{1}{2l+1}\int
_{0}^{\infty}db_{n}\ b_{n}\ \left[ 1-\left\vert S_{n}\left(
b_{n}\right) \right\vert ^{2}\right]  \
\sum_{m}\int_{0}^{\infty}d\rho\ \rho\ \int
_{0}^{2\pi}d\phi\ \nonumber\\
&  \times& \left\vert \int_{-\infty}^{\infty}dz\ \exp\left[ -ik_{z
}z\right]  \ \psi_{lm}\left(  \mathbf{r}\right) \right\vert ^{2}\
\left\vert
S_{c}\left(  b_{c}\right)  \right\vert ^{2}\ \nonumber\\
&  =& \frac{1}{2l+1}\int_{0}^{\infty}db_{n}\ b_{n}\ \left[
1-\left\vert
S_{n}\left(  b_{n}\right)  \right\vert ^{2}\right]  \ \sum_{m}C_{lm}^{2}%
\ \int_{0}^{\infty}d\rho\ \rho\nonumber\\
&  \times&\left\vert \mathcal{Z}_{lm}\left(  k_z,\rho\right)
\right\vert
^{2}\ \mathcal{S}\left(  b_{n},\rho\right)  , \label{siglong2}%
\end{eqnarray}
where%
\begin{eqnarray}
\mathcal{Z}_{lm}\left(  k_z,\rho\right)   &
=&\int_{-\infty}^{\infty }dz\ \exp\left[  -ik_zz\right]  \
R_{l}\left(  r\right)  \ P_{lm}\left(
\cos\theta\right)  \ \nonumber\\
\mathcal{S}\left(  b_{n},\rho\right)   & =&\int_{0}^{2\pi}d\phi\
\left\vert S_{c}\left( \sqrt{\rho^{2}+b_{n}^{2}-2\rho\
b_{n}\cos\phi}\right) \right\vert ^{2},
\end{eqnarray}
which is the same as eq. (\ref{strL}).


\begin{acknowledgments}
This work was supported by the U.\,S.\ National Science Foundation under grant
Nos.\ PHY-01\,10253 and PHY-00\,70911.
\end{acknowledgments}

%
%

\end{document}